\documentclass[%
 reprint,
preprint,
onecolumn,11pt,
amsmath,amssymb,aps,%prl,
prb,
]{revtex4-1}

\usepackage{verbatim}
\usepackage{graphicx}
\normalfont \topmargin -0.5 cm
\begin{document}

\title{Few-second-long correlation times in a quantum dot nuclear spin bath probed by frequency-comb NMR spectroscopy}

\author{A.~M.~Waeber$^1$}
%\email{e.chekhovich@sheffield.ac.uk}
\author{M.~Hopkinson$^2$}
\author{I.~Farrer$^3$}
\author{D.~A.~Ritchie$^3$}
\author{J.~Nilsson$^3$}
\author{R.~M.~Stevenson$^4$}
\author{A.~J.~Bennett$^4$}
\author{A.~J.~Shields$^4$}
\author{G.~Burkard$^5$}
\author{A.~I.~Tartakovskii$^1$}
\author{M.~S.~Skolnick$^1$}
\author{E.~A.~Chekhovich$^1$}
\affiliation{$^1$Department of Physics and Astronomy, University
of Sheffield, Sheffield S3 7RH, UK} \affiliation{$^2$Department of
Electronic and Electrical Engineering, University of Sheffield,
Sheffield S1 3JD, UK} \affiliation{$^3$Cavendish Laboratory,
University of Cambridge, CB3 0HE, UK}\affiliation{$^4$Toshiba
Research Europe Limited, Cambridge Research Laboratory, CB4 0GZ,
UK}\affiliation{$^5$Department of Physics, University of Konstanz,
D-78457 Konstanz, Germany}

\date{\today}

%\begin{abstract}
%Abstract
%\end{abstract}

%\pacs{73.21.La, 75.75.+a, 78.55.Et}% PACS, the Physics and Astronomy
                             % Classification Scheme.
%\keywords{Suggested keywords}%Use showkeys class option if keyword
                              %display desired
\maketitle

\textbf{One of the key challenges in spectroscopy is inhomogeneous
broadening that masks the homogeneous spectral lineshape and the
underlying coherent dynamics. A variety of techniques including
four-wave mixing and spectral hole-burning are used in optical
spectroscopy \cite{QDotFWM,VolkerHoleBurning,TwoColorNoiseSpec}
while in nuclear magnetic resonance (NMR) spin-echo
\cite{HahnEcho} is the most common way to counteract
inhomogeneity. However, the high-power pulses used in spin-echo
and other sequences
\cite{HahnEcho,NatureDynDecoupling,BathSuppression,InstDiffusion,PulsedSpinLocking}
often create spurious dynamics
\cite{InstDiffusion,PulsedSpinLocking} obscuring the subtle spin
correlations that play a crucial role in quantum information
applications
\cite{NatureDynDecoupling,BathSuppression,Merkulov,DasSarma2003,Sham2006,Bluhm,PressEcho,DeGreveHole,GreilichHoleQbit,HansomEnvAssisted,RMPReview}.
Here we develop NMR techniques that allow the correlation times of
the fluctuations in a nuclear spin bath of individual quantum dots
to be probed. This is achieved with the use of frequency comb
excitation which allows the homogeneous NMR lineshapes to be
measured avoiding high-power pulses. We find nuclear spin
correlation times exceeding 1 s in self-assembled InGaAs quantum
dots - four orders of magnitude longer than in strain-free III-V
semiconductors. The observed freezing of the nuclear spin
fluctuations opens the way for the design of quantum dot spin
qubits with a well-understood, highly stable nuclear spin bath.}

Pulsed magnetic resonance is a diverse toolkit with applications
in chemistry, biology and physics. In quantum information
applications, solid state spin qubits are of great interest and
are often described by the so called central spin model, where the
qubit (central spin) is coupled to a fluctuating spin bath
(typically interacting nuclear spins). Here microwave and
radio-frequency (rf) magnetic resonance pulses are used for the
initialization and readout of a qubit \cite{QSimulator}, dynamic
decoupling \cite{NatureDynDecoupling} and dynamic control
\cite{BathSuppression} of the spin bath.

However, the most important parameter controlling the central spin
coherence \cite{DasSarma2002,Merkulov,Sham2006} - the correlation
time $\tau_\textrm{c}$ of the spin bath fluctuations is very
difficult to measure directly. The $\tau_\textrm{c}$ is determined
by the spin exchange (flip-flops) of the interacting nuclear bath
spins. By contrast pulsed NMR reveals the spin bath coherence time
$T_2$, which characterizes the dynamics of the transverse nuclear
magnetization \cite{InstDiffusion,PulsedSpinLocking,QEchoNComms}
and is much shorter than $\tau_\textrm{c}$. The problem is further
exacerbated in self-assembled quantum dots where quadrupolar
effects lead to inhomogeneous NMR broadening exceeding 10~MHz
(Refs. \cite{QNMRNatNano,MunschNMR}), making rf field amplitudes
required for pulsed NMR practically unattainable.

Here we develop an alternative approach to NMR spectroscopy: we
measure non-coherent depolarisation of nuclear spins under weak
noise-like rf fields. Contrary to intuitive expectation, we show
that such measurement can reveal the full homogeneous NMR
lineshape describing the coherent spin dynamics. This is achieved
when rf excitation has a frequency comb profile (widely used in
precision optical metrology \cite{NatureFreqComb}). We then
exploit non-resonant nuclear-nuclear interactions: the homogeneous
NMR lineshape of one isotope measured with frequency comb NMR is
used as a sensitive non-invasive probe of the correlation times
$\tau_\textrm{c}$ of the nuclear flip-flops of the other isotope.
While initial studies \cite{DasSarma2002,Merkulov,RMPReview}
suggested $\tau_\textrm{c}\sim100~\mu$s for nuclear spins in III-V
semiconductors, it was recently recognized
\cite{KorenevQ,QEchoNComms,InGaAsNucDyn} that quadrupolar effects
may have a significant impact in self-assembled quantum dots. Here
we for the first time obtain a quantitative measurement of
extremely long $\tau_\textrm{c}\gtrsim1~$s revealing strong
freezing of the nuclear spin bath - a crucial advantage for
quantum information applications of self-assembled quantum dots.

\begin{figure*}
\includegraphics[bb=59pt 29pt 526pt 183pt, width=180mm]{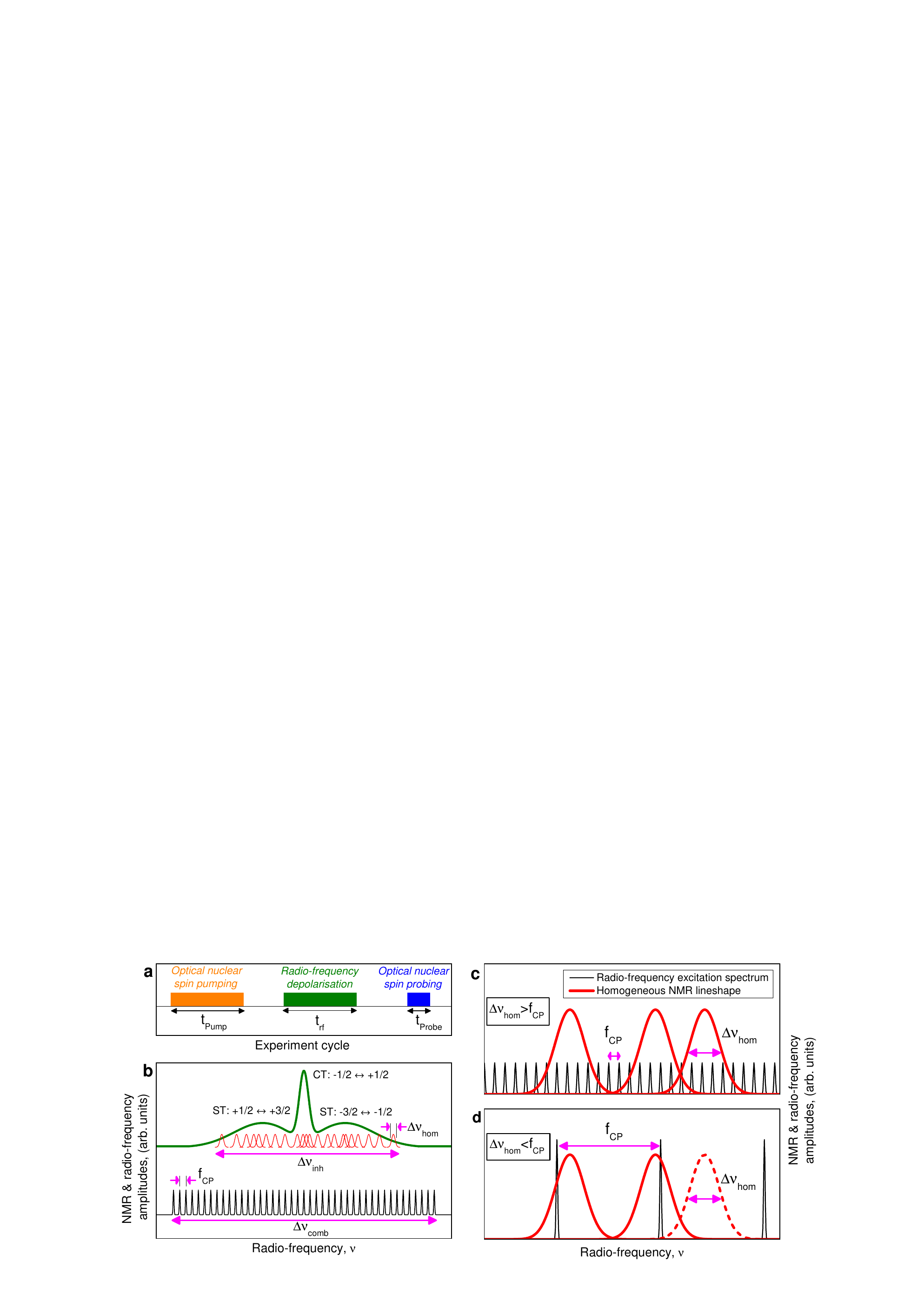}
\caption{\label{fig:MSScheme}\textbf{Frequency-comb technique for
homogeneous NMR lineshape measurement.} \textbf{a,} Timing diagram
of the experimental cycle consisting of nuclear spin optical
pumping (duration $t_\textrm{Pump}$), frequency-comb rf excitation
($t_\textrm{rf}$), and optical probing of the the nuclear spin
state ($t_\textrm{Probe}$). \textbf{b,} The green line shows
schematically a quantum dot NMR spectrum consisting of a central
transition (CT) peak and two satellite transition (ST) bands. The
inhomogeneous lineshape (width $\Delta\nu_{\textrm{inh}}$) is a
sum of a large number ($>$1000) of nuclear spin transitions with
homogeneous linewidths $\Delta\nu_{\textrm{hom}}$ (shown with red
lines). The thin black line shows the spectrum of the
frequency-comb excitation with comb period $f_\textrm{CP}$ and
total width $\Delta\nu_\textrm{comb}$ exceeding
$\Delta\nu_\textrm{inh}$. \textbf{c,} and \textbf{d,} demonstrate
how experiments with varying $f_\textrm{CP}$ can reveal the width
$\Delta\nu_{\textrm{hom}}$ of the NMR homogeneous lineshape (red
lines). When $\Delta\nu_{\textrm{hom}}>f_{\textrm{CP}}$
(\textbf{c}) all individual nuclei are uniformly excited by the
frequency comb (shown with black lines). In the opposite case
$\Delta\nu_{\textrm{hom}}<f_{\textrm{CP}}$ (\textbf{d}) some of
the nuclei (dashed line) are not excited resulting in a slow-down
of nuclear spin dynamics. The transition between the two cases
takes place when $f_\textrm{CP}\sim\Delta\nu_{\textrm{hom}}$,
allowing $\Delta\nu_{\textrm{hom}}$ to be measured.}
\end{figure*}

The experiments were performed on individual neutral
self-assembled InGaAs/GaAs quantum dots at magnetic field
$B_\textrm{z}=8$~T. All measurements of the nuclear spin
depolarisation dynamics employ the pump-depolarise-probe protocol
shown in Fig. \ref{fig:MSScheme}a. Here we exploit the hyperfine
interaction of the nuclei with the optically excited electron
\cite{RMPReview,QNMRNatNano,MunschNMR} both to polarise the nuclei
(pump pulse) and to measure the nuclear spin polarisation in terms
of the Overhauser shift $\Delta E_\textrm{hf}$ in the QD
photoluminescence spectrum (probe pulse). The rf magnetic field
depolarising nuclear spins is induced by a small copper coil.
(Further experimental details can be found in Methods and
Supplementary Note 1.)

All isotopes in the studied dots possess non-zero quadrupolar
moments. Here we focus on the spin $I=3/2$ nuclei $^{71}$Ga and
$^{75}$As. The strain-induced quadrupolar shifts result in an
inhomogeneously broadened NMR spectrum
\cite{QNMRNatNano,MunschNMR} as shown schematically by the green
line in Fig. \ref{fig:MSScheme}b. The spectrum consists of a
central transition (CT) $-1/2 \leftrightarrow +1/2$ and two
satellite transition (ST) $\pm1/2 \leftrightarrow \pm3/2$ peaks.
The NMR spectrum with inhomogeneous linewidth
$\Delta\nu_\textrm{inh}$ consists of individual nuclear spin
transitions (shown with red lines) with much smaller homogeneous
linewidth $\Delta\nu_\textrm{hom}$.

To make a non-coherent depolarisation experiment sensitive to the
homogeneous NMR lineshape rf excitation with a frequency comb
spectral profile is used. As shown in Fig. \ref{fig:MSScheme}b
(black line) the frequency comb has a period of $f_\textrm{CP}$
and a total comb width $\Delta\nu_\textrm{comb}$ exceeding
$\Delta\nu_\textrm{inh}$. The key idea of the frequency comb
technique is described in Figs. \ref{fig:MSScheme}c and d where
two possible cases are shown. If the comb period is small
($f_\textrm{CP}<\Delta\nu_\textrm{hom}$, Fig. \ref{fig:MSScheme}c)
all nuclear transitions are excited by a large number of rf modes.
As a result all nuclear spins are depolarised at the same rate and
we expect an exponential decay of the total nuclear spin
polarisation. In the opposite case of large comb period
($f_\textrm{CP}>\Delta\nu_\textrm{hom}$, Fig. \ref{fig:MSScheme}d)
some of the nuclear transitions are out of resonance and are not
excited (e.g. the one shown by the dashed red line). As a result
we expect a slowed-down non-exponential nuclear depolarisation.
The experiments are performed at different $f_\textrm{CP}$; the
$f_\textrm{CP}$ for which a slow-down in depolarisation is
observed gives a measure of the homogeneous linewidth
$\nu_\textrm{hom}$.

\begin{figure}
\includegraphics[bb=33pt 68pt 316pt 225pt, width=150mm]{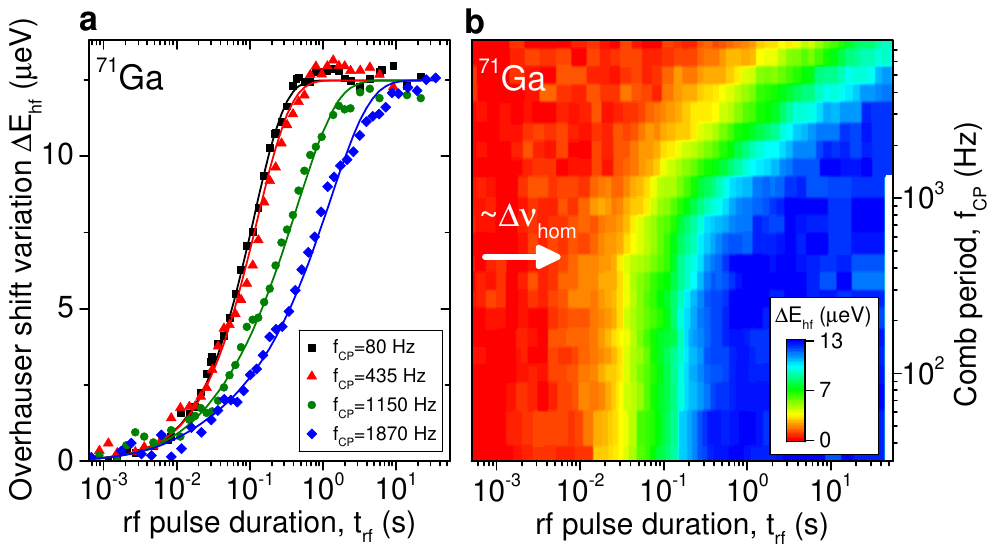}%82mm
\caption{\label{fig:MSResults} \textbf{Measurement of the
homogeneous NMR lineshape in self-assembled quantum dots using
frequency-comb excitation.} \textbf{a,} The change in the
polarisation of the $^{71}$Ga nuclear spins (measured in terms of
the change in the Overhauser shift $\Delta E_{\textrm{hf}}$) is
shown as a function of the rf pulse duration $t_{\textrm{rf}}$
(symbols) at $B_\textrm{z}=8$~T and different comb periods
$f_{\textrm{CP}}$. Lines show model fitting (see text).
\textbf{b,} A full 2D plot of $\Delta E_{\textrm{hf}}$ as a
function of $t_{\textrm{RF}}$ and $f_{\textrm{CP}}$ in the same
experiment as in \textbf{a}. There is a clear slow-down of the
nuclear spin depolarisation at $f_{\textrm{CP}}>450$~Hz (shown
with a white arrow) providing an estimate of the homogeneous
linewidth of $\Delta\nu_{\textrm{hom}}$.}
\end{figure}

Experimental demonstration of this technique is shown in Fig.
\ref{fig:MSResults}a. The Overhauser shift variation $\Delta
E_\textrm{hf}$ of $^{71}$Ga is shown as a function of the
depolarising rf pulse duration $t_\textrm{rf}$ for different
$f_\textrm{CP}$. For small $f_\textrm{CP}=$80 and 435~Hz an
exponential depolarisation is observed. However, when
$f_\textrm{CP}$ is increased the depolarisation becomes
non-exponential and slows down dramatically. The detailed
dependence $\Delta E_{\textrm{hf}}(t_\textrm{rf},f_\textrm{CP})$
is shown as a colour-coded plot in Fig. \ref{fig:MSResults}b. The
threshold value of $f_\textrm{CP}$ (marked with a white arrow)
above which the nuclear spin dynamics becomes sensitive to the
discrete structure of the frequency comb, provides an estimate of
$\Delta\nu_\textrm{hom}\sim$ 450~Hz. Such a small homogeneous
linewidth is detected in NMR resonances with inhomogeneous
broadening of $\Delta\nu_\textrm{inh}\sim6$~MHz (Ref.
\cite{QNMRNatNano}) demonstrating the resolution power of
frequency-comb non-coherent spectroscopy.

The information revealed by frequency-comb spectroscopy is not
limited to linewidth estimates. An accurate determination of the
full homogeneous lineshape is achieved with modeling based on
solving an integral equation (see details in Methods and
Supplementary Note 2). We use the following two-parameter
phenomenological model for the homogeneous lineshape:
\begin{eqnarray}
\begin{aligned}
&L(\nu)\propto\left(1+4(\sqrt[k]{2}-1)\frac{\nu^2}{\Delta\nu_\textrm{hom}^2}\right)^{-k},
\label{eq:kLineshape}
\end{aligned}
\end{eqnarray}
where $\Delta\nu_\textrm{hom}$ is the homogeneous full width at
half maximum and $k$ is a roll-off parameter that controls the
tails of the lineshape (the behavior of $L(\nu)$ at large $\nu$).
For $k=1$ the lineshape corresponds to Lorentzian, while for
$k\rightarrow\infty$ it tends to Gaussian: in this way Eq.
\ref{eq:kLineshape} seamlessly describes the two most common
lineshapes. Using $\Delta\nu_\textrm{hom}$ and $k$ as parameters
we calculate the model $\Delta
E_{\textrm{hf}}(t_\textrm{rf},f_\textrm{CP})$ dependence and fit
it to the experimental $\Delta
E_{\textrm{hf}}(t_\textrm{rf},f_\textrm{CP})$ to find an accurate
phenomenological description of the homogeneous NMR lineshape in
self-assembled quantum dots.

\begin{figure}
\includegraphics[bb=85pt 209pt 355pt 428pt, width=150mm]{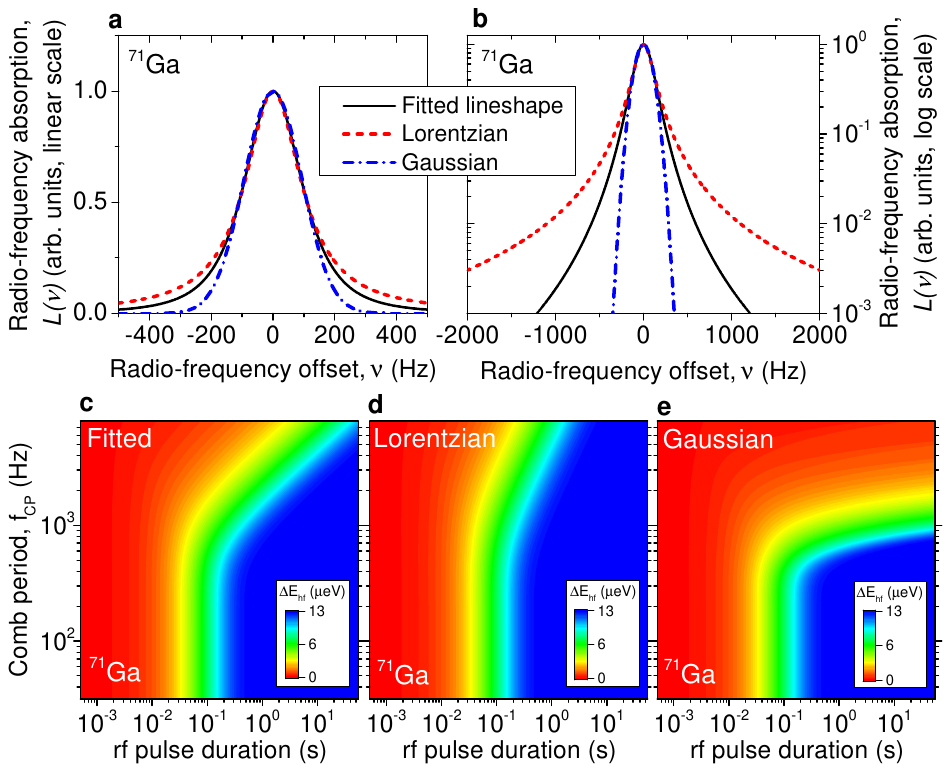}
\caption{\label{fig:MSModel}\textbf{Homogeneous lineshape
modeling.} \textbf{a, b,} Model homogeneous NMR lineshapes of
$^{71}$Ga nuclei shown on linear (\textbf{a}) and logarithmic
(\textbf{b}) scale. Solid lines show the best-fit lineshape with a
full width at half maximum $\Delta\nu_{\textrm{hom}}\approx221$~Hz
and a roll-off parameter $k\approx1.67$. Dashed and dashed-dotted
lines show for comparison Lorentzian and Gaussian lineshapes with
the same $\Delta\nu_{\textrm{hom}}$. \textbf{c-e,} The calculated
$\Delta E_{\textrm{hf}}(t_{\textrm{rf}},f_{\textrm{CP}})$
dependencies for the lineshapes in (\textbf{a, b}). For the fitted
lineshape (\textbf{c}) an excellent agreement with the experiment
in Fig. \ref{fig:MSResults}b is found, while calculations with
Lorentzian (\textbf{d}) and Gaussian (\textbf{e}) lineshapes give
markedly different results demonstrating the sensitivity of the
frequency comb technique.}
\end{figure}

The solid line in Fig. \ref{fig:MSModel}a shows the best-fit
lineshape ($\Delta\nu_\textrm{hom}\approx221$~Hz and
$k\approx1.67$) for the measurement shown in Figs.
\ref{fig:MSResults}a, b. The dashed and dashed-dotted lines in
Fig. \ref{fig:MSModel}a show for comparison the Lorentzian ($k=1$)
and Gaussian ($k\rightarrow\infty$) lineshapes with the same
$\Delta\nu_\textrm{hom}$. The difference in the lineshape tails is
seen clearly in Fig. \ref{fig:MSModel}b where a logarithmic scale
is used. The model $\Delta
E_{\textrm{hf}}(t_\textrm{rf},f_\textrm{CP})$ dependence
calculated with the best fit parameters is shown in Fig.
\ref{fig:MSModel}c and with lines in Fig. \ref{fig:MSResults}a -
there is excellent agreement with experiment. By contrast
modelling $\Delta E_{\textrm{hf}}(t_\textrm{rf},f_\textrm{CP})$
with Lorentzian (Fig. \ref{fig:MSModel}d) and Gaussian (Fig.
\ref{fig:MSModel}e) lineshapes show a pronounced deviation from
the experiment, demonstrating the excellent sensitivity of the
frequency-comb spectroscopy to accurately probe the homogeneous
spectral lineshape.

\begin{figure}
\includegraphics[bb=144pt 60pt 444pt 400pt, width=150mm]{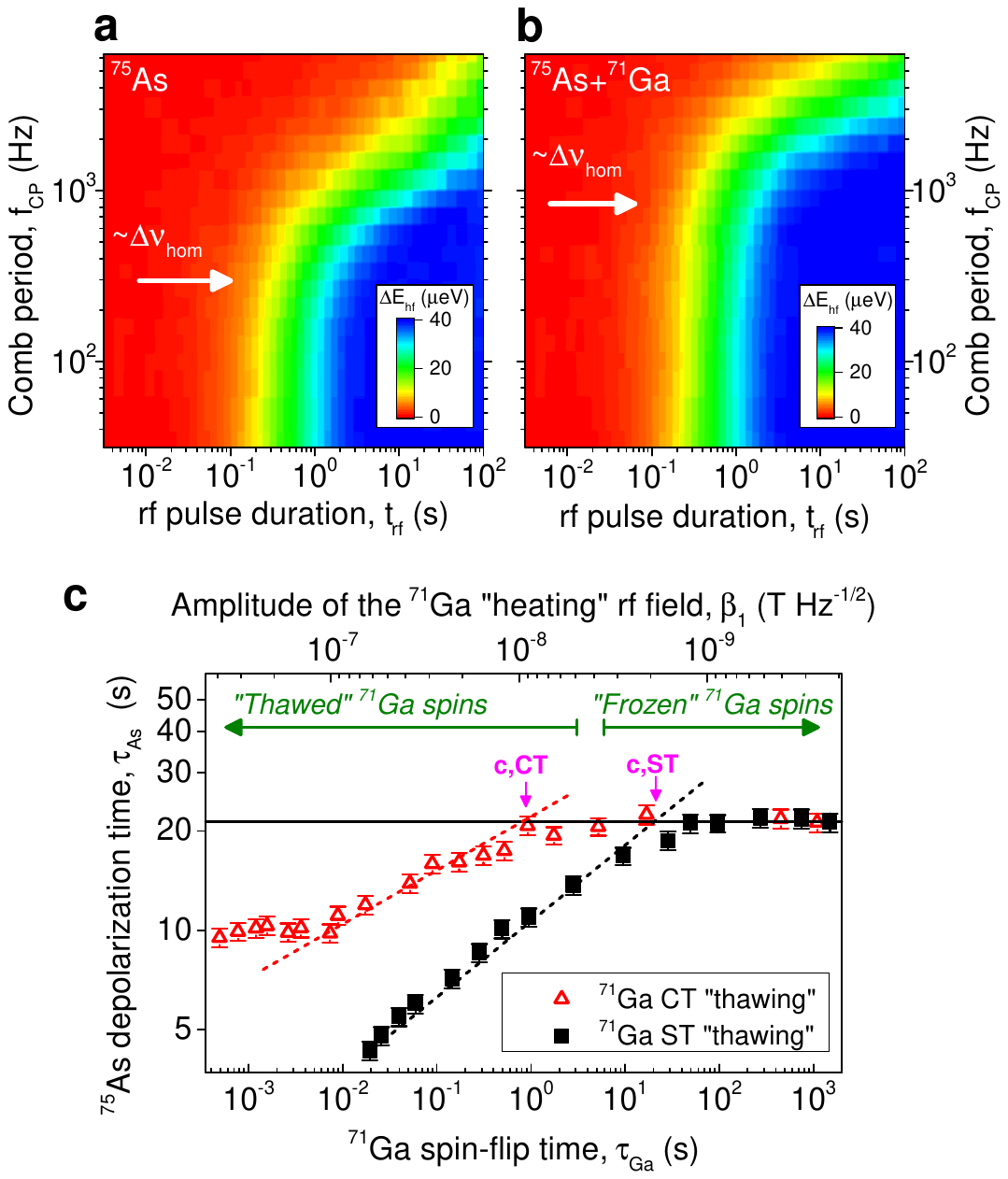}%78mm
\caption{\label{fig:MSTwoComb}\textbf{Probing the correlation
times of the nuclear spin bath fluctuation.} \textbf{a, b,}
Frequency-comb measurement on $^{75}$As nuclei at
$B_\textrm{z}=$8~T without (\textbf{a}) and with (\textbf{b})
additional frequency-comb excitation ''heating'' the $^{71}$Ga
nuclei. The increase of $\Delta\nu_\textrm{hom}$ of $^{75}$As
under the $^{71}$Ga ''heating'' reveals the strongly suppressed
nuclear spin fluctuations of $^{71}$Ga. \textbf{c,} rf-induced
depolarisation time $\tau_\textrm{As}$ of $^{75}$As nuclei at
fixed $f_\textrm{CP}=1.47$~kHz as a function of the amplitude
$\beta_1$ of the addition rf excitation ''heating'' either the
central transition (CT, triangles) or the satellite transition
(ST, squares) of $^{71}$Ga. The bottom scale shows $\beta_1$
expressed in terms of the rf-induced spin-flip time
$\tau_\textrm{Ga}$ of $^{71}$Ga. The $\tau_{\textrm{Ga}}$ at which
a marked decrease of $\tau_{\textrm{As}}$ is observed corresponds
to the correlation time $\tau_\textrm{c}$ of the $^{71}$Ga nuclear
spin fluctuations: By extrapolating the power-law regions (dashed
lines) to the intersection with the maximum $\tau_\textrm{As}$
(solid line) we deduce $\tau_\textrm{c,CT}\approx1$~sec and
$\tau_\textrm{c,ST}\approx20$~sec for the CT and ST transitions.}
\end{figure}

We have also performed frequency comb NMR spectroscopy on
$^{75}$As nuclei (Fig. \ref{fig:MSTwoComb}a). Despite their larger
inhomogeneous broadening $\Delta\nu_\textrm{inh}\sim18$~MHz the
model fitting reveals even smaller
$\Delta\nu_{\textrm{hom}}\approx117$~Hz and $k\approx1.78$. The
frequency-comb measurements are in agreement with the previous
findings based on spin-echo NMR measurements \cite{QEchoNComms}:
indeed, from $\Delta\nu_{\textrm{hom}}$ derived here we can
estimate the nuclear spin coherence time
$T_2\approx1/(\pi\Delta\nu_\textrm{hom})\sim1.4$ and $2.7$~ms for
$^{71}$Ga and $^{75}$As, in good agreement with the corresponding
spin-echo $T_2\approx1.2$ and $4.3~$ms. On the other hand,
spin-echo could only be measured on central transitions for which
$\Delta\nu_{\textrm{inh}}$ is relatively small. Moreover pulsed
NMR does not allow determination of the full homogeneous
lineshape, which for dipole-dipole interactions typically has a
''top-hat''-like (Guassian) profile \cite{VanVleck}. And most
importantly, due to the parasitic effects such as ''instantaneous
diffusion'' \cite{InstDiffusion} and spin locking
\cite{PulsedSpinLocking} pulsed NMR does not reveal the
characteristic correlation time $\tau_\textrm{c}$ of the spin
exchange (spin flip-flop) between the nuclei in the absence of rf
excitation.

As we now show, the non-Gaussian lineshapes can be understood and
$\tau_\textrm{c}$ can be derived using experiments with two
frequency combs exciting nuclei of two isotopes ($^{75}$As and
$^{71}$Ga). The two-comb experiment is similar to that shown in
Fig. \ref{fig:MSTwoComb}a: we excite $^{75}$As nuclei with a
frequency comb to measure their homogeneous lineshape. The
difference is that now we simultaneously apply a second comb
exciting the $^{71}$Ga spins. Importantly, in this experiment the
$^{71}$Ga nuclei are first fully depolarised after the optical
nuclear spin pumping -- in this way the excitation of $^{71}$Ga
has no direct effect on the measured hyperfine shift $\Delta
E_{\textrm{hf}}$. By contrast it leads to ''heating'' of the
$^{71}$Ga spins which has only an indirect effect on $\Delta
E_{\textrm{hf}}$ by changing the $^{75}$As lineshape via dipolar
coupling between $^{71}$Ga and $^{75}$As spins. The result of the
two-comb experiment is shown in Fig. \ref{fig:MSTwoComb}b: a clear
increase of $\Delta\nu_{\textrm{hom}}$ for $^{75}$As is observed.
From model fitting we find that $^{71}$Ga ''heating'' leads to a 3
times broader homogeneous linewidth
$\Delta\nu_{\textrm{hom}}\approx355$~Hz of $^{75}$As and its
homogeneous lineshape is modified towards Gaussian, observed as
increased $k\approx2.32$.

To explain this result we note that the NMR lineshape is a
statistical distribution of NMR frequency shifts of each nucleus
produced by its dipolar interaction with all possible
configurations of the neighboring nuclear spins. However, the
frequency comb experiment is limited in time (up to $\sim$100~s as
shown in Figs. \ref{fig:MSTwoComb}a, b). If the nuclear spin
environment of each $^{75}$As nucleus does not go through all
possible configurations during the measurement time, the frequency
shifts are effectively static, and hence are eliminated from the
lineshape as for any other inhomogeneous broadening.

Thus we conclude that the narrowed, non-Gaussian
($k\approx$1.6--1.8) homogeneous NMR lineshape arises from the
''snapshot'' nature of the frequency comb measurement, probing the
strongly frozen nuclear spin configuration. When the additional
$^{71}$Ga ''heating'' excitation is applied it ''thaws'' the
$^{71}$Ga spins, detected as broadening of the $^{75}$As lineshape
(as demonstrated in Figs. \ref{fig:MSTwoComb}a, b). We use such
sensitivity of the $^{75}$As lineshape to measure the dynamics of
the $^{71}$Ga equilibrium spin bath fluctuations. Based on the
results of Fig. \ref{fig:MSTwoComb} the $^{75}$As spins are now
excited with a frequency comb with a fixed
$f_\textrm{CP}=1.47$~kHz for which the $^{75}$As depolarisation
dynamics is most sensitive to the $^{71}$Ga ''heating''.
Furthermore, we now use selective ''heating'' of either the
$-1/2\leftrightarrow+1/2$ CT or the $+1/2\leftrightarrow+3/2$ ST
of $^{71}$Ga. The amplitude $\beta_1$ of the ''heating'' frequency
comb is varied -- the resulting dependencies of the $^{75}$As
depolarisation time $\tau_{\textrm{As}}$ are shown in Fig.
\ref{fig:MSTwoComb}c by the squares and triangles for CT and ST
''heating'' respectively. For analysis we also express $\beta_1$
in terms of the rf-induced $^{71}$Ga spin-flip time
$\tau_{\textrm{Ga}}$ (bottom scale in Fig. \ref{fig:MSTwoComb}c,
see details in Methods).

It can be seen that for vanishing $^{71}$Ga excitation
($\beta\rightarrow0$) the $^{75}$As depolarisation time
$\tau_{\textrm{As}}$ is constant. In this ''frozen'' regime the
rf-induced spin-flip time $\tau_{\textrm{Ga}}$ of $^{71}$Ga is
larger than the correlation time $\tau_\textrm{c}$ of the
$^{71}$Ga intrinsic spin flip-flops
($\tau_{\textrm{Ga}}>\tau_\textrm{c}$). As a result
$\tau_{\textrm{As}}$ is determined only by the rf excitation of
$^{75}$As itself. However, when $\beta_1$ is increased to
$\sim$1--10~nT~Hz$^{-1/2}$ the rf induced spin-flips of $^{71}$Ga
nuclei become faster than their intrinsic flip-flops
($\tau_{\textrm{Ga}}<\tau_\textrm{c}$). Such ''thawing'' of
$^{71}$Ga broadens the $^{75}$As lineshape (via heteronuclear
interaction), and is observed as a reduction of
$\tau_{\textrm{As}}$. Thus the transition from the ''frozen'' to
''thawed'' regimes takes place when
$\tau_{\textrm{Ga}}\approx\tau_\textrm{c}$, allowing
$\tau_\textrm{c}$ to be determined. In Fig. \ref{fig:MSTwoComb}c
we extrapolate graphically (dashed lines) the power-law dependence
in the ''thawed'' regime. The points of the intersections with the
limiting value of $\tau_{\textrm{As}}$ in the ''frozen'' regime
(solid horizontal line) yield correlation times
$\tau_\textrm{c,CT}\sim1$~s for CT and
$\tau_\textrm{c,ST}\sim20$~s for ST.

The observed $\tau_\textrm{c}\gtrsim1$~s exceeds very strongly
typical nuclear dipolar flip-flop times in strain-free III-V
solids $\tau_\textrm{c}\sim100~\mu$s
\cite{DasSarma2003,Merkulov,RMPReview}. We attribute the extremely
long $\tau_\textrm{c}$ in self-assembled quantum dots to the
effect of inhomogeneous nuclear quadrupolar shifts making nuclear
spin flip-flops energetically forbidden
\cite{KorenevQ,QEchoNComms}. This interpretation is corroborated
by the observation of $\tau_\textrm{c,ST}\gg\tau_\textrm{c,CT}$,
since quadrupolar broadening of the ST transitions is much larger
than that of the CT \cite{QNMRNatNano}. Furthermore, the $^{71}$Ga
spins examined here have the largest gyromagnetic ratio $\gamma$
and the smallest quadrupolar moment $Q$, so we expect that all
other isotopes in InGaAs have even longer $\tau_\textrm{c}$,
resulting in the overall $\tau_\textrm{c}\gtrsim1$~s of the entire
quantum dot nuclear spin bath. This implies that in high magnetic
fields the spin-echo coherence times of the electron and hole spin
qubits in self-assembled dots are not limited by the nuclear spin
bath up to sub-second regimes
\cite{DasSarma2003,Khaetskii,Merkulov,Bluhm}. Provided that other
mechanisms of central spin dephasing, such as charge fluctuations
\cite{PressEcho,DeGreveHole,GreilichHoleQbit} are eliminated, this
would open the way for optically active spin qubit networks in
III-V semiconductors with coherence properties previously
achievable only in nuclear-spin-free materials
\cite{MuhonenSiQubit,NeumannDiamond}.

Since the frequency-comb technique is not limited by artifacts in
the spin dynamics hampering pulsed magnetic resonance, it allows
detection of very slow spin bath fluctuations. Such sensitivity of
the method can be used for example to investigate directly the
effect of the electron or hole on the spin bath fluctuations in
charged quantum dots, arising for example from hyperfine-mediated
nuclear spin interactions. The experiments can be well understood
within a classical rate equation model, while further advances in
frequency comb spectroscopy can be expected with the development
of a full quantum mechanical model. Furthermore, the simple and
powerful ideas of frequency-comb NMR spectroscopy can be readily
extended beyond quantum dots: as we show in Supplementary Note 3
the only essential requirement is that the longitudinal relaxation
time $T_1$ should be larger (by about two orders of magnitude)
than the transverse relaxation time $T_2$, which is usually the
case in solid state spin systems. Finally our approaches in the
use of frequency combs can go beyond NMR, and for example enrich
the techniques in optical spectroscopy.

\section{\label{sec:Methods} Methods Summary}

\textbf{Sample structures and experimental techniques} The
experiments were performed on individual neutral self-assembled
InGaAs/GaAs quantum dots. The sample was mounted in a helium-bath
cryostat ($T$=4.2~K) with a magnetic field $B_\textrm{z}=8$~T
applied in the Faraday configuration (along the sample growth and
light propagation direction $Oz$). Radio-frequency (rf) magnetic
field $B_\textrm{rf}$ perpendicular to $B_\textrm{z}$ was induced
by a miniature copper coil. Optical excitation was used to induce
nuclear spin magnetization exceeding 50\%, as well as to probe it
by measuring hyperfine shifts in photoluminescence
spectroscopy\cite{QNMRNatNano}.

Two sample structures have been studied, both containing a single
layer of InGaAs/GaAs quantum dots embedded in a weak planar
microcavity with a Q-factor of $\sim$250. In one of the samples
the dots emitting at $\sim945$~nm were placed in a $p-i-n$
structure, where application of a large reverse bias during the rf
excitation ensured the neutral state of the dots. The results for
this sample are shown in Fig. \ref{fig:MSResults}. The second
sample was a gate-free structure, where most of the dots emitting
at $\sim914$~nm are found in a neutral state, although the
charging can not be controlled. Excellent agreement between the
lineshapes of both $^{71}$Ga and $^{75}$As in the two structures
was found, confirming the reproducibility of the frequency-comb
technique.

\textbf{Homogeneous lineshape theoretical model.} Let us consider
an ensemble of spin $I=1/2$ nuclei with gyromagnetic ratio
$\gamma$ and inhomogeneously broadened distribution of nuclear
resonant frequencies $\nu_\textrm{nuc}$. We assume that each
nucleus has a homogeneous absorption lineshape $L(\nu)$, with
normalization $\int_{-\infty}^{+\infty}L(\nu)d\nu=1$.
%i.e. L has units of [s]
A small amplitude (non-saturating) rf field will result in
depolarisation, which can be described by a differential equation
for population probabilities $p_{\pm1/2}$ of the nuclear spin
levels $I_\textrm{z}=\pm1/2$
\begin{eqnarray}
d(p_{+1/2}-p_{-1/2})/dt=-W(p_{+1/2}-p_{-1/2}).\label{eq:RFDepODE}
\end{eqnarray}
For frequency-comb excitation the decay rate is the sum of the
decay rates caused by each rf mode with magnetic field amplitude
$B_\textrm{1}$, and can be written as:
\begin{eqnarray}
W(\nu_\textrm{nuc})=\frac{\gamma^2B_\textrm{1}^2}{2f_\textrm{CP}}\sum_{j=0}^{N_\textrm{m}-1}
L(\nu_\textrm{nuc}-\nu_1-j
f_\textrm{CP})f_\textrm{CP},\label{eq:RFDepRate}
\end{eqnarray}
where the summation goes over all modes with frequencies
$\nu_\textrm{j}=\nu_\textrm{1}+j f_\textrm{CP}$ ($\nu_\textrm{1}$
is the frequency of the first spectral mode).

The change in the Overhauser shift $E_{\textrm{hf}}$ produced by
each nucleus is proportional to $p_{+1/2}-p_{-1/2}$ and according
to Eq. \ref{eq:RFDepODE} has an exponential time dependence
$\propto \exp(-W(\nu_\textrm{nuc})t)$. The quantum dot contains a
large number of nuclear spins with randomly distributed absorption
frequencies. Therefore to obtain the dynamics of the total
Overhauser shift we need to average over $\nu_\textrm{nuc}$, which
can be done over one period $f_\textrm{CP}$ since the spectrum of
the rf excitation is periodic. Furthermore, since the total width
of the rf frequency comb $\Delta\nu_\textrm{comb}$ is much larger
than $f_\textrm{CP}$ and $\Delta\nu_\textrm{hom}$, the summation
in Eq. \ref{eq:RFDepRate} can be extended to $\pm\infty$. Thus,
the following expression is obtained for the time dependence
$\Delta E_{\textrm{hf}}(t,f_\textrm{CP})$, describing the dynamics
of the rf-induced nuclear spin depolarisation:
\begin{eqnarray}
\begin{aligned}
&\frac{\Delta E_{\textrm{hf}}(t,f_\textrm{CP})}{\Delta
E_\textrm{hf}(t\rightarrow\infty)}=1-
%\\
%&
f_\textrm{CP}^{-1}\int\limits_0^{f_\textrm{CP}}\exp\left(-t
\frac{\gamma^2B_\textrm{1}^2}{2f_\textrm{CP}}
\sum_{j=-\infty}^{\infty} L(\nu_\textrm{nuc}-j
f_\textrm{CP})f_\textrm{CP}\right)d\nu_\textrm{nuc}.
\label{eq:MSRFDec}
\end{aligned}
\end{eqnarray}

Equation \ref{eq:MSRFDec} describes the dependence $\Delta
E_{\textrm{hf}}(t,f_\textrm{CP})$ directly measurable in
experiments such as shown in Fig. \ref{fig:MSResults}b. $\Delta
E_\textrm{hf}(t\rightarrow\infty)$ is the total optically induced
Overhauser shift of the studied isotope and is also measurable,
while $f_\textrm{CP}$ and $B_\textrm{1}$ are parameters that are
controlled in the experiment. We note that in the limit of small
comb period $f_\textrm{CP}\rightarrow0$ the infinite sum in Eq.
\ref{eq:MSRFDec} tends to the integral
$\int_{-\infty}^{+\infty}L(\nu)d\nu=1$ and the Overhauser shift
decay is exponential (as observed experimentally) with a
characteristic time
\begin{eqnarray}
\begin{aligned}
\tau=2f_\textrm{CP}/(\gamma^2B_\textrm{1}^2) \label{eq:ExpDecRate}
\end{aligned}
\end{eqnarray}

Equation \ref{eq:MSRFDec} is a Fredholm's integral equation of the
first kind on the homogeneous lineshape function $L(\nu)$. This is
an ill-conditioned problem: as a result finding the lineshape
requires some constraints to be placed on $L(\nu)$. Our approach
is to use a model lineshape of Eq. \ref{eq:kLineshape}. After
substituting $L(\nu)$ from Eq. \ref{eq:kLineshape}, the right-hand
side of Eq. \ref{eq:MSRFDec} becomes a function of the parameters
$\Delta\nu_\textrm{hom}$ and $k$ which we then find by
least-squares fitting of Eq. \ref{eq:MSRFDec} to the experimental
dependence $\Delta E_{\textrm{hf}}(t,f_\textrm{CP})$.

This model is readily extended to the case of $I>1/2$ nuclei. Eq.
\ref{eq:RFDepODE} becomes a tri-diagonal system of differential
equations, and the solution (Eq. \ref{eq:RFDepRate}) contains a
sum of multiple exponents under the integral. These modifications
are straightforward but tedious and can be found in Supplementary
Note 2.

\textbf{Derivation of the nuclear spin bath correlation times.}
Accurate lineshape modeling is crucial in revealing the $^{75}$As
homogeneous broadening arising from $^{71}$Ga ''heating''
excitation (as demonstrated in Figs. \ref{fig:MSTwoComb}a, b).
However, since a measurement of the full $\Delta
E_{\textrm{hf}}(t,f_\textrm{CP})$ dependence is time consuming,
the experiments with variable $^{71}$Ga excitation amplitude
$\beta_1$ (Fig. \ref{fig:MSTwoComb}c) were conducted at fixed
$f_\textrm{CP}=1.47$~kHz exceeding noticeably the $^{75}$As
homogeneous linewidth $\Delta\nu_{\textrm{hom}}\approx117$~Hz. To
extract the arsenic depolarisation time $\tau_\textrm{As}$ we fit
the arsenic depolarisation dynamics $\Delta E_{\textrm{hf}}(t)$
with the following formulae: $\Delta
E_{\textrm{hf}}(t_\textrm{rf})=\Delta
E_\textrm{hf}(t_\textrm{rf}\rightarrow\infty)(1-\exp[-(t_\textrm{rf}/\tau_\textrm{As})^r])$,
using $r$ as a common fitting parameter and $\tau_\textrm{As}$
independent for measurements with different $\tau_\textrm{Ga}$. We
find $r\approx0.57$, while the dependence $\tau_\textrm{As}$ on
$\tau_\textrm{Ga}$ obtained from the fit is shown in Fig.
\ref{fig:MSTwoComb}c with error bars corresponding to 95\%
confidence intervals.

The period of the $^{71}$Ga ''heating'' frequency comb is kept at
a small value $f_\textrm{CP}=150$~Hz ensuring uniform excitation
of all nuclear spin transitions. The amplitude of the ''heating''
comb is defined as $\beta_1=B_1/\sqrt{f_\textrm{CP}}$, where $B_1$
is magnetic field amplitude of each mode in the comb (further
details can be found in Supplementary Note 1). To determine the
correlation times we express $\beta_1$ in terms of the rf-induced
spin-flip time $\tau_{\textrm{Ga}}$. The $\tau_{\textrm{Ga}}$ is
defined as the exponential time of the $^{71}$Ga depolarisation
induced by the ''heating'' comb and is derived from an additional
calibration measurement. The values of $\beta_1$ shown in Fig.
\ref{fig:MSTwoComb}c correspond to the experiment on the CT and
are calculated using Eq. \ref{eq:ExpDecRate} as
$\sqrt{2/(4\gamma^2\tau_{\textrm{Ga}})}$, where $\gamma$ is the
$^{71}$Ga gyromagnetic ratio and $\tau_{\textrm{Ga}}$ is
experimentally measured. The additional factor of 4 in the
denominator is due to the matrix element of the CT of spin
$I=3/2$. For experiments on ST the $\beta_1$ values shown in Fig.
\ref{fig:MSTwoComb}c must be multiplied by $\sqrt{4/3}$.

\textbf{ACKNOWLEDGMENTS} The authors are grateful to K.V. Kavokin
for useful discussions. This work has been supported by the EPSRC
Programme Grant EP/J007544/1, ITN S$^3$NANO. E.A.C. was supported
by a University of Sheffield Vice-Chancellor's Fellowship. I.F.
and D.A.R. were supported by EPSRC.

%\textbf{AUTHOR CONTRIBUTIONS} .

\textbf{ADDITIONAL INFORMATION} Correspondence and requests for
materials should be addressed to A.M.W (a.waeber@sheffield.ac.uk)
or E.A.C. (e.chekhovich@sheffield.ac.uk).

%\bibliography{bibl}
%\end{document}

\renewcommand{\thesection}{Supplementary Note \arabic{section}}
\setcounter{section}{0}
\renewcommand{\thefigure}{\arabic{figure}}
\renewcommand{\figurename}{Supplementary Figure}
\setcounter{figure}{0}
\renewcommand{\theequation}{\arabic{equation}}
\setcounter{equation}{0}
\renewcommand{\thetable}{Supplementary Table \arabic{table}}
\setcounter{table}{0}

\renewcommand{\citenumfont}[1]{S#1}
\makeatletter
\renewcommand{\@biblabel}[1]{S#1.}
\makeatother

\pagebreak \pagenumbering{arabic}

\section*{SUPPLEMENTARY INFORMATION}

\begin{figure}[b]
\includegraphics[bb=90pt 53pt 440pt 370pt]{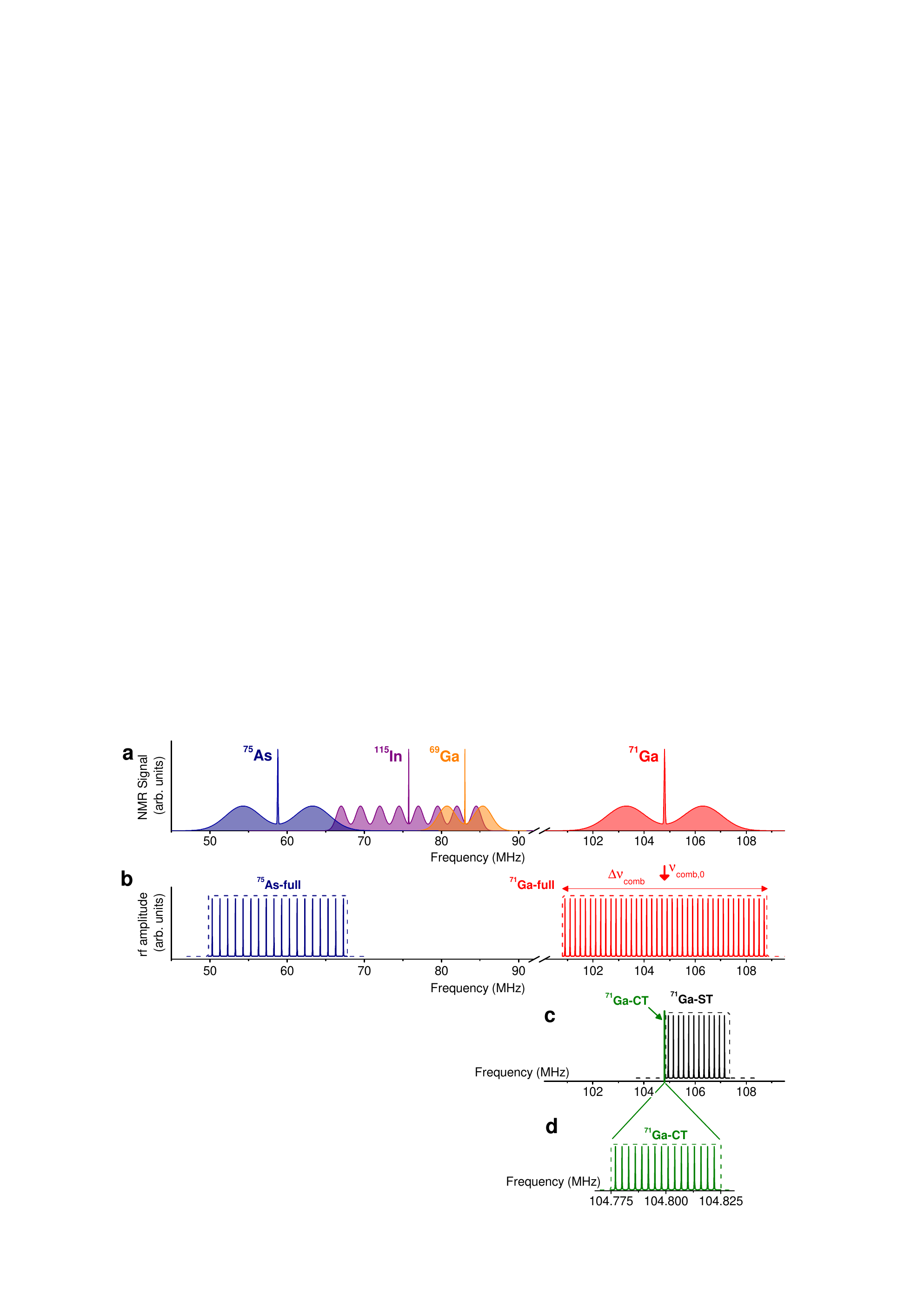}
\caption{\label{fig:FreqCombRF} \textbf{Frequency combs for NMR
experiments.} \textbf{a,} Schematic NMR spectrum of a strained
self-assembled InGaAs quantum dot at $B_z=8$~T. Sharp peaks arise
from the central transitions (CTs) between $-1/2$ and $+1/2$
nuclear spin levels of each isotope. There are two satellite
transitions (STs) for the spin-3/2 isotopes of Ga and As, and
eight STs for the spin-9/2 indium. Inhomogeneous quadrupolar
broadening results in significant overlap of the $^{115}$In and
$^{69}$Ga inhomogeneous lineshapes, while the overlap of $^{75}$As
and $^{115}$In is minimal. The $^{71}$Ga resonance is well
isolated. \textbf{b-d,} Schematic spectra of the frequency combs
used in experiments. The $^{75}$As-full comb is used to excite the
entire inhomogeneous lineshape of $^{75}$As with minimum effect on
$^{115}$In. Similarly $^{71}$Ga-full comb excites the entire
$^{71}$Ga resonance. Frequency comb $^{71}$Ga-ST is used to excite
selectively the $+1/2\leftrightarrow+3/2$ ST of $^{71}$Ga, while
the comb $^{71}$Ga-CT is used to excite the
$-1/2\leftrightarrow+1/2$ CT (zoom in is shown in \textbf{d}). The
parameters of all frequency combs are summarized in
\ref{tab:FreqCombPars}, further details can be found in
\ref{SI:TechniquesFreqComb}.}
\end{figure}

\begin{figure}
\includegraphics[bb=83pt 58pt 423pt 266pt]{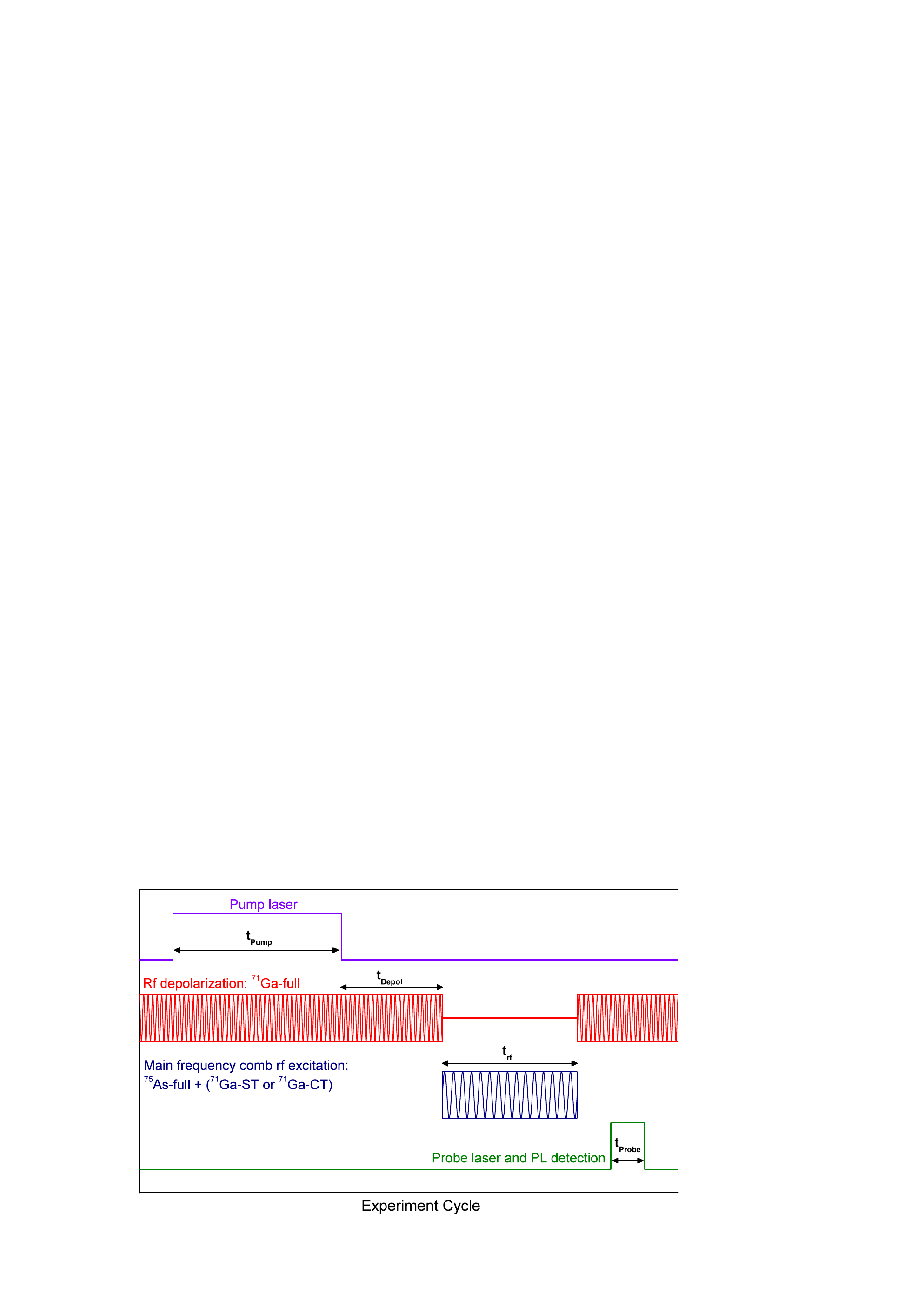}
\caption{\label{fig:TDiagrS} \textbf{Time diagram of an optically
detected frequency-comb nuclear magnetic resonance experiment.}
See detailed explanation in \ref{SI:TechniquesTDiagr}.}
\end{figure}

\begin{figure}
\includegraphics[bb=12pt 80pt 550pt 273pt, width=190mm]{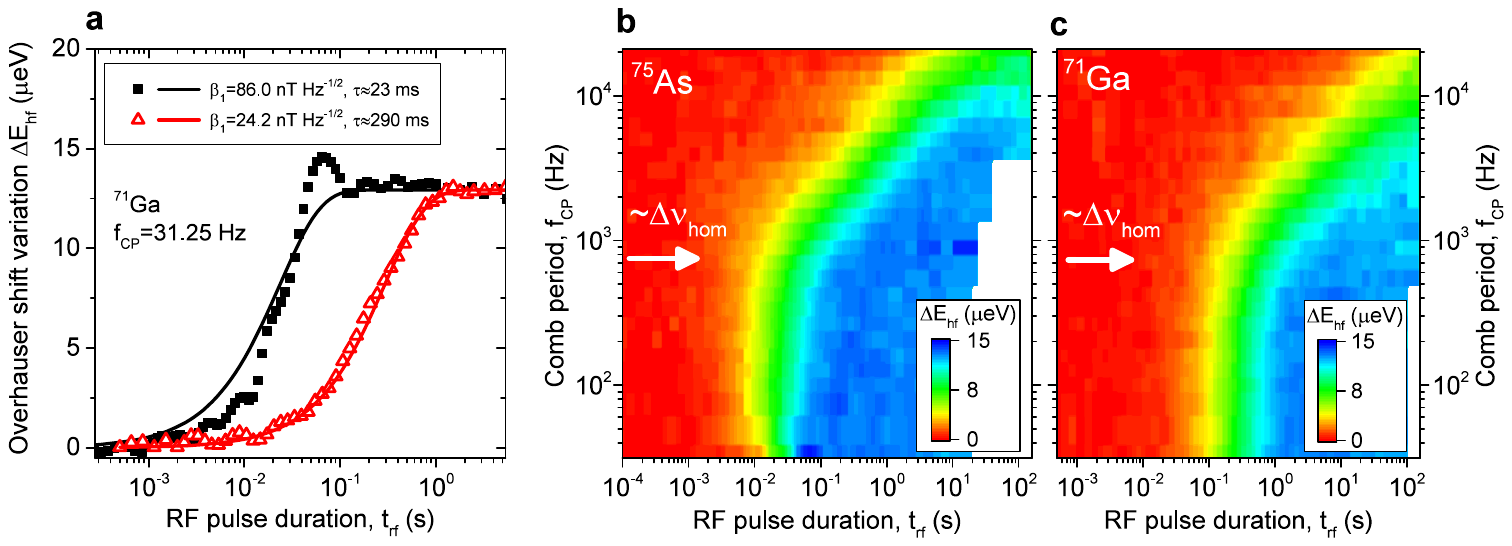}
\caption{\label{fig:RFPowerDep} \textbf{Experimental verification
of the rate equation model.} \textbf{a,} Dynamics of the $^{71}$Ga
nuclear spin depolarisation at $B_z=8$~T induced by frequency comb
rf excitation with comb period $f_\textrm{CP}=31.25$~Hz. The
results at high rf field density $\beta_1=86.0$~nT Hz$^{-1/2}$
(low rf field density $\beta_1=24.2$~nT Hz$^{-1/2}$) are shown by
the squares (triangles). Solid lines show exponential fits with
decay time $\tau\approx23$~ms ($\tau\approx290$~ms) for high (low)
rf field density. \textbf{b, c,} Full $\Delta
E_{\textrm{hf}}(t,f_\textrm{CP})$ dependence measured at high
(\textbf{b}) and low rf (\textbf{c}) field densities.}
\end{figure}

\pagebreak

\begin{table}
\caption{\label{tab:FreqCombPars} Frequency comb parameters used
in NMR experiments.}
\begin{ruledtabular}
\begin{tabular}{lccccc}  % here I choose if it is centred or left etc
\multicolumn{5}{l}{\textbf{a,} Homogeneous lineshape experiments (Fig. 2):}\\
\hline
Isotope & $^{75}$As & $^{71}$Ga &  &  \\
\hline
Frequency comb& $^{75}$As-full & $^{71}$Ga-full &  &  \\
\hline
Comb central frequency $\nu_\textrm{comb,0}$ (MHz) & 58.81 & 104.80 &  &  \\
\hline
Comb spectral width $\Delta\nu_\textrm{comb}$ (MHz) & 18 & 9 &  &  \\
\hline
Comb period $f_\textrm{CP}$ (Hz)  & varied  & varied &  &  \\
\hline
Rf field density $\beta_1$ ($\textrm{nT}\ \textrm{Hz}^{-1/2}$) & 65.5 & 39.1 &  &  \\
\hline\hline
\multicolumn{5}{l}{\textbf{b,} Lineshape broadening experiments (Figs. 4a, 4b):}\\
\hline
Isotope & $^{75}$As & $^{71}$Ga &  &  \\
\hline
Frequency comb& $^{75}$As-full & $^{71}$Ga-full &  &  \\
\hline
Comb central frequency $\nu_\textrm{comb,0}$ (MHz) & 58.81 & 104.80 &  &  \\
\hline
Comb spectral width $\Delta\nu_\textrm{comb}$ (MHz) & 18 & 8 &  &  \\
\hline
Comb period $f_\textrm{CP}$ (Hz)  & varied & 159 &  &  \\
\hline
Rf field density $\beta_1$ ($\textrm{nT}\ \textrm{Hz}^{-1/2}$) & 30.5 & 27.6 &  &  \\
\hline\hline
\multicolumn{5}{l}{\textbf{c,} Correlation time experiments (Fig. 4c):}\\
\hline
Isotope & $^{75}$As & $^{71}$Ga & $^{71}$Ga & $^{71}$Ga \\
\hline
Frequency comb& $^{75}$As-full & $^{71}$Ga-full & $^{71}$Ga-ST & $^{71}$Ga-CT \\
\hline
Comb central frequency $\nu_\textrm{comb,0}$ (MHz) & 58.81 & 104.80 & 106.10 & 104.80 \\
\hline
Comb spectral width $\Delta\nu_\textrm{comb}$ (MHz) & 18 & 8 & 2.5 & 0.05 \\
\hline
Comb period $f_\textrm{CP}$ (Hz)  & 1466 & 159 & 150 & 150 \\
\hline
Rf field density $\beta_1$ ($\textrm{nT}\ \textrm{Hz}^{-1/2}$) & 30.5 & 27.6 & varied & varied \\
\end{tabular}
\end{ruledtabular}
\end{table}

\clearpage

\section{\label{SI:Techniques}Details of experimental techniques}

\subsection{\label{SI:TechniquesFreqComb}Frequency combs}

The key novel findings of this work are based on the use of
radiofrequency (rf) excitation with a frequency comb spectral
profile. Here we give detailed parameters of the frequency combs
used in the experiments on self-assembled quantum dots.

Supplementary Figure \ref{fig:FreqCombRF}a shows a schematic NMR
spectrum of a self-assembled InGaAs quantum dot at $B_z=8$~T,
based on results of the inverse NMR measurements
\cite{QNMRNatNano}. Four sharp peaks arise from the central
transitions (CTs) $-1/2\leftrightarrow+1/2$ of each isotope. All
spin-3/2 nuclei ($^{75}$As, $^{69}$Ga and $^{71}$Ga) have two
satellite transitions (STs) $-3/2\leftrightarrow-1/2$ and
$+1/2\leftrightarrow+3/2$ observed as inhomogeneously broadened
bands on both sides of the CTs. The spin-9/2 $^{115}$In nuclei
have a total of eight STs. For clarity these are shown as four
bands on each side of the CT, although in experimental NMR spectra
the peaks arising from different STs merge forming two
inhomogeneously broadened bands on both sides of the $^{115}$In CT
\cite{QNMRNatNano}. As shown in Supplementary Figure
\ref{fig:FreqCombRF}a the spectral contributions from the
$^{115}$In and $^{69}$Ga nuclei overlap significantly;
furthermore, there is a small overlap between the $^{75}$As and
$^{115}$In NMR resonances.

In the experiments we use rf frequency combs that are designed to
influence only the chosen transition(s) of one isotope without
affecting the other isotopes as demonstrated in Supplementary
Figures \ref{fig:FreqCombRF}b-d. The frequency-comb labeled
$^{71}$Ga-full has a total width $\Delta\nu_\textrm{comb}$ of 8 or
9~MHz. This comb entirely covers the inhomogeneous lineshape of
$^{71}$Ga and thus uniformly excites all nuclear spin transitions
of this isotope. Since $^{71}$Ga has a large resonance frequency
the $^{71}$Ga-full comb does not affect the polarisation of the
other isotopes. To achieve selective excitation of the $^{71}$Ga
ST ($+1/2\leftrightarrow+3/2$) we use frequency comb $^{71}$Ga-ST
with $\Delta\nu_\textrm{comb}=2.5$~MHz (Supplementary Fig.
\ref{fig:FreqCombRF}c). Similarly, selective excitation of
$^{71}$Ga CT is achieved with a narrow comb $^{71}$Ga-CT with a
width of $\Delta\nu_\textrm{comb}=50$~kHz as shown in
Supplementary Figs. \ref{fig:FreqCombRF}c, d.

In the case of $^{75}$As we use a frequency comb $^{75}$As-full
that excites the entire inhomogeneouse resonance line. This comb
inevitably excites some of the $^{115}$In nuclear transitions,
mostly $-9/2\leftrightarrow-7/2$ STs. However, excitation of the
$-9/2\leftrightarrow-7/2$ ST alone has a negligible effect on the
overall change in the nuclear polarisation of the spin-9/2 nuclei
of $^{115}$In. Thus the $^{75}$As-full comb with the optimum width
$\Delta\nu_\textrm{comb}=18$~MHz is used in experiments for
selective depolarisation of $^{75}$As.

The central frequencies $\nu_\textrm{comb,0}$, the widths
$\Delta\nu_\textrm{comb}$ and the comb periods (spectral
separation between the adjacent modes) $f_\textrm{CP}$ of all the
combs used in experiments are listed in \ref{tab:FreqCombPars}.
Depending on the experiment the $f_\textrm{CP}$ is either varied
or kept constant. Typically the values of $f_\textrm{CP}$ ranging
from 30~Hz to 21~kHz are employed, so that the total number of
modes in the comb
$N_\textrm{m}=\Delta\nu_\textrm{comb}/f_\textrm{CP}+1$ ranges from
$\sim$330 to $\sim$600000.

The phases of individual modes of the frequency comb are chosen in
a way that minimizes the peak power for a given average power
(i.e. a waveform with the minimum crest-factor). In particular the
following expression satisfies this criterion:
\begin{equation}
B_\textrm{rf}(t)=B_{1}\sum _{j=1}^{N_\textrm{m}} \cos \left( 2 \pi
\left(\nu_1+(j-1) f_{\textrm{CP}}\right)t+\pi\frac{j
(j-1)}{N_\textrm{m}}\right)\label{eq:FreqCombSignal},
\end{equation}
where the summation goes over all modes, and $\nu_1$ is the
frequency of the first mode of the comb. In experiments the
frequency-comb signal is generated by an arbitrary waveform
generator equipped with a 64 million points memory.

Each mode of the frequency comb has the same amplitude $B_1$ of
the rf oscillating magnetic field. In experiments where the comb
period $f_\textrm{CP}$ is varied $B_1$ has to be adjusted to
maintain the same total power of the frequency comb excitation.
Since the power is proportional to $B_1^2$ the
$B_1^2/f_\textrm{CP}$ ratio has to be kept constant. Thus the
amplitude of the frequency comb can be conveniently characterized
by the magnetic field density
$\beta_\textrm{1}=B_1/\sqrt{f_\textrm{CP}}$.

The amplitude of the frequency comb rf magnetic field can be
calibrated from an additional pulsed Rabi oscillation experiment
on the CT of a selected isotope \cite{QEchoNComms}. The Rabi
oscillation circular frequency $\omega_\textrm{Rabi}$ can be
measured experimentally and is proportional to the rf field
amplitude $B_\textrm{rf}$ in the rotating frame:
\begin{equation}
B_\textrm{rf}=\frac{\omega_\textrm{Rabi}}{2
\gamma}\label{eq:BRabi},
\end{equation}
where $\gamma$ is nuclear gyromagnetic ratio and the factor $2$
originates from the dipole transition matrix element of the CT of
a spin-3/2 nucleus \cite{AbragamBook}. Throughout all experiments,
we monitored the applied rf fields via a pick-up coil that was
placed close to the sample and connected to a spectrum analyzer.
By comparing the voltages induced by the frequency comb modes with
the voltage associated with a field $B_\textrm{rf}$ in the Rabi
oscillation experiment, we derived the values of $B_1$ and
$\beta_\textrm{1}$ of the comb. The values of $\beta_\textrm{1}$
used in different experiments are given in \ref{tab:FreqCombPars}
- these correspond to the rotating frame, i.e. the physical values
of the magnetic field induced by the coil are twice as large.

\subsection{\label{SI:TechniquesTDiagr}Optical pump-probe techniques for frequency-comb NMR}

Supplementary Fig. \ref{fig:TDiagrS} shows the time sequence of
optical and rf excitation pulses used in a frequency comb
measurement of the equilibrium nuclear spin bath fluctuations of
the $^{71}$Ga isotope (the results are shown in Fig. 4c of the
main text). As explained in the main text this experiment is based
on measuring the rf induced dynamics of the $^{75}$As spins in the
presence of additional rf excitation of $^{71}$Ga spins. The
experimental cycle consists of the following four stages described
below.

\textit{Optical nuclear spin pumping.} At the start of each new
measurement cycle, the nuclear spin bath is reinitialized
optically. This is achieved with optically induced dynamic nuclear
polarisation (DNP) \cite{Eble, LowPowDNP}: under high power,
circularly polarised laser excitation, spin polarised electrons
are created. These electrons can efficiently transfer their
polarisation to the nuclear spin bath via hyperfine interaction
\cite{GammonPRL}. We use $\sigma^-$ polarised excitation with a
pump laser operating at $\sim$850~nm, in resonance with the QD
wetting layer. By using sufficiently long pumping times
($t_\textrm{Pump}=5.5-6.5$~s) and high powers $\sim 10
P_\textrm{sat}$ (where $P_\textrm{sat}$ is the saturation power of
the QD ground states), we create a reproducible and high degree of
nuclear polarisation.

\textit{Depolarisation of $^{71}$Ga isotope.} Optical spin pumping
polarises the nuclei of all isotopes. However, in order to probe
the equilibrium fluctuations of $^{71}$Ga spins their longitudinal
relaxation has to be excluded from the measured dynamics. For that
$^{71}$Ga nuclear polarisation has to be erased, which is achieved
by exciting the spins with a $^{71}$Ga-full frequency comb for a
sufficiently long time $t_\textrm{Depol}=1.2$~s. To simplify
experimental implementation the depolarising rf is kept on during
the nuclear spin pumping stage as well, which has no effect on the
experimental results.

\textit{Frequency comb rf excitation.} Following the spin bath
preparation (optical DNP and $^{71}$Ga depolarisation), the main
rf excitation (variable duration $t_\textrm{rf}$) is applied. For
the spin bath fluctuation measurement (Fig. 4c) this excitation is
a sum of the $^{75}$As-full frequency comb and either the
$^{71}$Ga-ST or the $^{71}$Ga-CT comb. Since $^{71}$Ga is
completely depolarised by the previous pulse, all changes in the
total nuclear polarisation at this stage are solely due to the
$^{75}$As depolarisation. In this way we ensure that it is the
depolarisation dynamics of $^{75}$As that is measured, while the
$^{71}$Ga-ST or $^{71}$Ga-CT ''heating'' excitation only induces
nuclear spin-flips of the corresponding $^{71}$Ga transition.

\textit{Optical probing of the nuclear spin polarisation.} At the
end of the experiment cycle, a short probe laser pulse is applied
and the resulting photoluminescence spectrum is collected by a 1~m
double spectrometer with a CCD. The changes in the quantum dot
Zeeman splitting (the Overhauser shift $E_\textrm{hf}$) are used
to probe the nuclear spin state. The probe laser is non-resonant
($\sim850$~nm), yet unlike the pump laser it is linearly polarised
and the probe power and duration $t_\textrm{Probe}$ are chosen
such that no noticeable DNP is induced and the final nuclear spin
polarisation is measured accurately. Typical probe parameters used
in the experiments were $t_\textrm{Probe}=4$~ms and $\sim
P_\textrm{sat}/10$ for QDs in the $p-i-n$ diode sample and
$t_\textrm{Probe}=60$~ms and $\sim P_\textrm{sat}/50$ in the
gate-free structure. Depending on the QD photoluminescence
intensity the experiment cycle was repeated $10-40$ times to
achieve the optimum signal-to-noise ratio.

For the purpose of data analysis we are interested in measuring
the rf-induced change in the nuclear spin polarisation (rf-induced
change in the Overhauser shift $\Delta E_{\textrm{hf}}$), for that
we perform a control measurement where $^{75}$As-full comb is off,
and subtract the resulting QD Zeeman splitting from the Zeeman
splittings obtained in the measurements with $^{75}$As excitation.
In this way $\Delta E_{\textrm{hf}}=0$ corresponds to no nuclear
spin depolarisation induced by the rf.

The diagram of Supplementary Fig. \ref{fig:TDiagrS} also describes
the other types of frequency comb NMR measurements presented in
the main text with the following modifications: For the line
broadening measurements shown in Figs. 4a, b the main excitation
is a sum of the $^{75}$As-full and $^{71}$Ga-full frequency combs
(the $^{71}$Ga-full is off for the measurement in Fig. 4a and is
on for Fig. 4b). The homogeneous lineshape measurement (Fig. 2) is
performed without the additional rf depolarisation excitation,
while for the main rf excitation the $^{71}$Ga-full comb is used.

\section{\label{SI:Model}Theoretical model for nuclear spin dynamics under frequency-comb excitation}

The Methods section of the main text describes the model for spin
$I=1/2$ nuclei. Here we consider a more general case of nuclear
spins $I>1/2$ with gyromagnetic ratio $\gamma$. We consider the
Overhauser shifts of nuclei of only one isotope, which is
justified since the polarisation of other isotopes stays constant
during the measurement and can be neglected. In an external
magnetic field the nuclear spin state is split into $2I+1$ states
with spin projections $I_\textrm{z}=-I,\ -I+1,\ ...\ I-1,\ +I$. In
our classical rate equation model we assume that each nuclear spin
has a probability $p_\textrm{m}$ to be found in a state with
$I_\textrm{z}=m$ with normalization condition
\begin{eqnarray}
\sum_{m=-I}^I p_{m}(t)=1.\label{eq:pNorm}
\end{eqnarray}
We also assume that at $t=0$ optical pumping initializes all
nuclear spins into a Boltzman distribution
\begin{eqnarray}
p_{m}(t=0)=p_{m,0}\propto\exp(\xi m),\label{eq:BoltzmanInit}
\end{eqnarray}
so that nuclear spins can be characterized by a temperature
$T_\textrm{nuc}\propto1/\xi$. When optical nuclear spin pumping is
used, $T_\textrm{nuc}$ is very small compared to the spin bath
temperature $T$, so the equilibrium nuclear polarisation can be
neglected.

Application of the radiofrequency (rf) excitation leads to the
changes in population probabilities. The rf excites only
dipole-allowed transitions for which $I_\textrm{z}$ changes by
$\pm1$. In the experiment we use weak, non-saturating
radiofrequency fields. Thus instead of the full Bloch equations
for nuclear magnetization, the evolution of the population
probabilities of the states with $I_\textrm{z}=m$ can be described
with the following first-order differential equation (see further
details in \ref{SI:TechniquesApplicability}):
\begin{eqnarray}
\begin{aligned}
&dp_{m}/dt=-(W_{m-1,m}+W_{m,m+1}) p_{m}(t)+W_{m-1,m} p_{m-1}(t)+W_{m,m+1} p_{m+1}(t),\ -I<m<I\\
&dp_{I}/dt=W_{I-1,I}(-p_{I}(t)+p_{I-1}(t))\\
&dp_{-I}/dt=W_{-I,-I+1}
(-p_{-I}(t)+p_{-I+1}(t)).\label{eq:RFDepODEs}
\end{aligned}
\end{eqnarray}
Here the first equation describes the states with $-I<I_z<I$. Its
first term is due to nuclei with $I_\textrm{z}=m$ making
transitions into the $I_\textrm{z}=m-1$ and $m+1$ states, whereas
the second and third terms describe the opposite case of nuclei
transitioning into the $I_\textrm{z}=m$ state from the
$I_\textrm{z}=m-1$ and $I_\textrm{z}=m+1$ states respectively. The
second and third equations correspond to the case of $m=-I$ and
$m=+I$ respectively. Taken for all $m$, for which $-I\leq m\leq
I$, the Supplementary Eq. \ref{eq:RFDepODEs} yields a system of
$2I+1$ first-order ordinary differential equations (ODEs) for
$2I+1$ time-dependent variables $p_{m}(t)$ with initial conditions
given by Supplementary Eq.~\ref{eq:BoltzmanInit}. Due to the
normalization condition of Supplementary Eq.~\ref{eq:pNorm}, only
$2I$ variables and equations are independent. This system of ODEs
has a tridiagonal matrix with coefficients determined by the rf
induced transition rates $W_{m,m+1}$ which satisfy a symmetry
condition $W_{m,m+1}=W_{m+1,m}$.

Similar to the case of $I=1/2$, each transition rate $W_{m,m+1}$
resulting from the frequency-comb excitation is a sum of
transition rates caused by individual rf modes each having
magnetic field amplitude $B_\textrm{1}$. We assume that each
nuclear transition has the same broadening described by the
homogeneous lineshape function $L(\nu)$, with normalization
$\int_{-\infty}^{+\infty}L(\nu)d\nu=1$. Due to the inhomogeneous
quadrupolar shifts the NMR transition frequency
$\nu_\textrm{m,m+1}$ is generally different for each pair of spin
levels $I_\textrm{z}=m$ and $I_\textrm{z}=m+1$ (even for one
nucleus). Thus for the transition rates we can write:
\begin{eqnarray}
W_{m,m+1}(\nu_{m,m+1})=\frac{(I-m)(I+m+1)\gamma^2B_\textrm{1}^2}{2f_\textrm{CP}}\sum_{j=-\infty}^{+\infty}
L(\nu_{m,m+1}-\nu_1-j
f_\textrm{CP})f_\textrm{CP},\label{eq:RFDepRates}
\end{eqnarray}
where the summation goes over all modes with frequencies
$\nu_\textrm{j}=\nu_\textrm{1}+(j-1) f_\textrm{CP}$, and is
extended to $\pm\infty$ since the total width of the rf excitation
comb $\Delta\nu_\textrm{comb}$ is much larger than $f_\textrm{CP}$
and the homogeneous linewidth $\Delta\nu_\textrm{hom}$. The
$(I-m)(I+m+1)$ factor arises from the dipolar transition matrix
element \cite{AbragamBook}. We note that Supplementary Eq.
\ref{eq:RFDepODEs} does not involve any explicit nuclear-nuclear
interactions. Instead such interactions are introduced in
Supplementary Eq. \ref{eq:RFDepRates} phenomenologically via the
homogeneous broadening described by the lineshape function
$L(\nu)$. On the other hand the presence of finite homogeneous
broadening is essential in order to use the limit of weak rf
fields \cite{BlochAnalyt} and transform the Bloch equations into
rate equations (Supplementary Eq. \ref{eq:RFDepODEs}). The
validity of the weak rf field approximation is discussed and
verified experimentally in \ref{SI:TechniquesApplicability}.

Since Supplementary Eq. \ref{eq:RFDepODEs} is a system of linear
first-order equations, the solution is a multiexponential
relaxation towards the fully depolarised state where all nuclear
spin states have equal populations $p_{m}=1/(2I+1)$. The solution
has the general form:
\begin{eqnarray}
p_{m}(t)=1/(2I+1)+\sum_\textrm{j=1}^{2I}a_{m,\textrm{j}}\exp(-\lambda_\textrm{j}
t),\label{eq:RFDepSols}
\end{eqnarray}
where $\lambda_\textrm{j}$ are the non-zero eigenvalues of the ODE
system matrix of Supplementary Eq. \ref{eq:RFDepODEs}. The values
of $\lambda_\textrm{j}$ depend on all transition rates $W_{m,m+1}$
from Supplementary Eq.~\ref{eq:RFDepRates}, while the coefficients
$a_{m,\textrm{j}}$ depend both on $W_{m,m+1}$ and the initial
probabilities $p_{m,0}$ from Supplementary
Eq.~\ref{eq:BoltzmanInit}.

Non-zero nuclear spin polarisation along the magnetic field ($Oz$
axis) changes the spectral splitting of the quantum dot Zeeman
doublet. Such change known as the Overhauser shift is measured
experimentally using photoluminescence spectroscopy. The time
evolution of the Overhauser shift for the fixed values of nuclear
transition frequencies $\{\nu_{m,m+1}\}=\{\nu_{-I,-I+1},\
\nu_{-I+1,-I+2},\ ...\ \nu_{I-1,I}\}$ reads as:
\begin{eqnarray}
E_{\textrm{hf,1}}(t,f_\textrm{CP},B_1,T_\textrm{nuc},L(\nu),\{\nu_{m,m+1}\})=A\sum_{m=-I}^{+I}m
p_{m}(t),\label{eq:OHSRates}
\end{eqnarray}
where $A$ is the hyperfine constant and we used Supplementary
Eqns.~\ref{eq:BoltzmanInit}, \ref{eq:RFDepRates},
\ref{eq:RFDepSols} so that $E_{\textrm{hf,1}}$ is dependent on
$f_\textrm{CP}$, $B_\textrm{1}$, $T_\textrm{nuc}$, the homogeneous
lineshape function $L(\nu)$ and all nuclear transition frequencies
$\{\nu_{m,m+1}\}$ as parameters.

Each QD contains a large number of nuclear spins with randomly
distributed absorption frequencies. Thus to describe the
experiment on nuclear spins in a self-assembled quantum dot we
need to average over all $\nu_\textrm{m,m+1}$, which can be done
over one period $f_\textrm{CP}$ since the spectrum of the
frequency-comb rf excitation is periodic. Similarly to the case of
$I=1/2$, the following expression is obtained for the time
dependence of the Overhauser shift, describing the dynamics of
rf-induced nuclear spin depolarisation:
\begin{eqnarray}
\begin{aligned}
&E_{\textrm{hf}}(t,f_\textrm{CP},B_1,T_\textrm{nuc},L(\nu))=\\
&f_\textrm{CP}^{-2I} \int\limits_0^{f_\textrm{CP}}d\nu_{-I,-I+1}\
...\int\limits_0^{f_\textrm{CP}}d\nu_{I-1,I}\
E_{\textrm{hf,1}}(t,f_\textrm{CP},B_1,T_\textrm{nuc},L(\nu),\{\nu_{m,m+1}\}).
\label{eq:MSRFDecs}
\end{aligned}
\end{eqnarray}

The quantity measured in the experiment is the rf-induced
variation of the Overhauser shift:
\begin{eqnarray}
\begin{aligned}
\Delta
E_{\textrm{hf}}(t,f_\textrm{CP},B_1,T_\textrm{nuc},L(\nu))=E_{\textrm{hf}}(t,f_\textrm{CP},B_1,T_\textrm{nuc},L(\nu))-E_{\textrm{hf}}(t=0).
\label{eq:MSRFDecDeltas}
\end{aligned}
\end{eqnarray}
The values of $f_\textrm{CP}$ and $B_\textrm{1}$ are the
parameters that are controlled in the experiment. The nuclear spin
temperature $T_\textrm{nuc}$ can be determined using the known
hyperfine constant $A$ and the measured total Overhauser shift
$\Delta E_{\textrm{hf}}(t=\infty)$. Thus for a given homogeneous
NMR lineshape $L(\nu)$ the nuclear spin depolarisation dynamics
can be fully predicted from Supplementary Eqns.
\ref{eq:pNorm}--\ref{eq:MSRFDecDeltas}. Conversely, Supplementary
Eq. \ref{eq:MSRFDecs} can be treated as an integral equation on
the unknown homogeneous lineshape function $L(\nu)$. Since
Fredholm's integral equation of the first kind is an
ill-conditioned problem, some constrains on $L(\nu)$ are required.
Our approach is to use a model lineshape with two parameters
$\Delta\nu_\textrm{hom}$ and $k$ (Eq. 1 of the main text). Upon
substituting this model lineshape the Overhauser shift variation
of Supplementary Eq. \ref{eq:MSRFDecDeltas} becomes $\Delta
E_{\textrm{hf}}(t,f_\textrm{CP},B_1,T_\textrm{nuc},\Delta\nu_\textrm{hom},k)$.
We then perform least-square fitting to the experimental
dependence $\Delta E_{\textrm{hf}}(t,f_\textrm{CP})$ using $B_1$,
the homogeneous linewidth $\Delta\nu_\textrm{hom}$ and the
roll-off parameter $k$ as fitting parameters and using the nuclear
spin temperature $T_\textrm{nuc}$ obtained from the experiment.

The ODE system of Supplementary Eq. \ref{eq:RFDepODEs} can be
solved analytically for $I\leq3/2$, however it turns out to be
more practical to perform numerical diagonalization in order to
obtain the eigenvalues $\lambda_\textrm{j}$ and coefficients
$a_{m,\textrm{j}}$ which are then used in Supplementary Eq.
\ref{eq:RFDepSols}. Similarly we use numerical integration to
evaluate Supplementary Eq. \ref{eq:MSRFDecs}.

\section{\label{SI:TechniquesApplicability}Applicability of the
frequency comb technique and the rate equation model.}

The evolution of the nuclear magnetization under rf excitation can
be described by the Bloch equations \cite{Bloch}. In this model
the solution under resonant monochromatic excitation is determined
by the three important parameters: the amplitude of the resonant
field $B_\textrm{rf}$, and the relaxation times characterizing the
system, the longitudinal $T_1$ and the transverse $T_2$. In
self-assembled quantum dots $T_1$ is extremely long (few hours
\cite{InGaAsNucDyn,InPDyn}), so that the longitudinal relaxation
can be neglected. Thus the nuclear spin dynamics is determined by
the relation between $B_\textrm{rf}$ and $T_2$. Two cases are
possible \cite{BlochAnalyt}. If rf magnetic field is strong
($\gamma B_\textrm{rf} T_2\gg 1$) the nuclear magnetization has
oscillatory behaviour (Rabi oscillations are observed). By
contrast, for weak rf excitation ($\gamma B_\textrm{rf} T_2\ll 1$)
there are no oscillations, and any nuclear magnetization $I_z$
along the external field decays exponentially to its steady state
value \cite{BlochAnalyt}. The exponential dynamics in the weak rf
excitation regime allow for the problem to be simplified and for
the rate equation model described by Supplementary Eqns.
\ref{eq:RFDepODEs} to be used. The validity of the rate equation
model is essential for the determination of the homogeneous
lineshape and thus sets the applicability limit for the frequency
comb technique itself.

To verify the applicability of the rate equation model we
performed frequency comb spectroscopy measurements at different rf
amplitudes. Supplementary figure \ref{fig:RFPowerDep} shows the
results for $^{71}$Ga measured at high rf field density
$\beta_1=86.0$~nT~Hz$^{-1/2}$ and low rf field density
$\beta_1=24.2$~nT~Hz$^{-1/2}$ -- in all other respects the
conditions in these experiments were the same as in the experiment
with medium $\beta_1=39.1$~nT Hz$^{-1/2}$ shown in Fig. 2 of the
main text.

The Supplementary figure \ref{fig:RFPowerDep}a shows by symbols
the nuclear spin depolarisation dynamics at small comb period
$f_\textrm{CP}=31.25$~Hz. At low rf field amplitude
$\beta_1=24.2$~nT~Hz$^{-1/2}$ (triangles) the decay can be
described very well by a single exponential decay
($\tau\approx290$~ms) shown with a solid line. By contrast, at
high amplitude $\beta_1=86.0$~nT~Hz$^{-1/2}$ (squares) the
dynamics shows signatures of oscillations, and there is a clear
deviation from the exponential behaviour (best exponential fit is
for $\tau\approx23$~ms).

Supplementary Figures \ref{fig:RFPowerDep}b and c show the full
$\Delta E_{\textrm{hf}}(t,f_\textrm{CP})$ dependencies measured at
high (\textbf{b}) and low rf (\textbf{c}) field densities. The
$\Delta E_{\textrm{hf}}(t,f_\textrm{CP})$ profiles are in good
agreement for the two experiments (except for the rescaling along
the $t_\textrm{rf}$ axis). Using model fitting we find
$\Delta\nu_\textrm{hom}\approx244$~Hz, $k\approx1.66$ for
$\beta_1=86.0$~nT Hz$^{-1/2}$ and
$\Delta\nu_\textrm{hom}\approx223$~Hz, $k\approx1.68$ for
$\beta_1=24.2$~nT Hz$^{-1/2}$. This is also in good agreement with
$\Delta\nu_\textrm{hom}\approx221$~Hz, $k\approx1.67$ found for
the measurement at $\beta_1=39.1$~nT Hz$^{-1/2}$ shown in Fig. 2
of the main text.

We thus conclude that the discrepancy between the results in
Supplementary Figs. \ref{fig:RFPowerDep}b and c becomes
significant only at small comb period $f_\textrm{CP}=31.25$~Hz (as
also demonstrated in Supplementary Figure \ref{fig:RFPowerDep}a).
This can be explained as follows: At high rf excitation amplitude
the nuclear spin depolarisation takes place on a shorter time
scale $\tau$. If the rf pulses are shorter than $1/f_\textrm{CP}$,
the spectral profile of the frequency comb becomes distorted. Thus
if the nuclear polarisation decay timescales $\tau$ are shorter
than $1/f_\textrm{CP}$, the rate equation model is no longer
applicable since the rf excitation can not be described as a
frequency comb. Thus it is required that $\tau>1/f_\textrm{CP}$.
Furthermore, to measure the homogeneous lineshape and linewidth
$\Delta\nu_\textrm{hom}$ we only need to use frequency combs with
comb periods $f_\textrm{CP}$ comparable to or larger than
$\Delta\nu_\textrm{hom}$, so it is required that
$f_\textrm{CP}\in\{\Delta\nu_\textrm{hom},\infty\}$. Combining
$\tau>1/f_\textrm{CP}$ and
$f_\textrm{CP}\in\{\Delta\nu_\textrm{hom},\infty\}$ we find the
following condition on the frequency comb technique applicability:
\begin{eqnarray}
\begin{aligned}
\Delta\nu_\textrm{hom}>1/\tau, \label{eq:FCCondition}
\end{aligned}
\end{eqnarray}
which restricts the rf amplitude, characterized by the
depolarisation time $\tau$.

Another requirement, arising from the applicability of the weak rf
field limit of the Bloch equations, is that the rf induced
depolarisation time $\tau$ must be longer than the transverse
relaxation time $T_2$. However, $T_2$ is related to the
homogeneous linewidth as $T_2\sim1/(\pi\Delta\nu_\textrm{hom})$.
Thus the requirement $\tau>T_2$ leads to the same condition as
that of Supplementary Eq. \ref{eq:FCCondition}. Furthermore, since
the frequency comb technique relies on the measurement of the
longitudinal nuclear magnetization, the depolarisation time $\tau$
must be shorter than the nuclear spin $T_1$ times. In combination
with Supplementary Eq. \ref{eq:FCCondition} this leads to the
following condition:
\begin{eqnarray}
\begin{aligned}
\Delta\nu_\textrm{hom}>1/\tau>1/T_1 \label{eq:FCCondition2}
\end{aligned}
\end{eqnarray}
This condition has a dual role: it sets the boundaries for the rf
excitation amplitude (characterized by $\tau$) and sets the
limitation $\Delta\nu_\textrm{hom}>1/T_1$ on the properties of the
nuclear spin system that can be studied with the frequency comb
technique. This latter condition can be rewritten as
\begin{eqnarray}
\begin{aligned}
T_1>a T_2 \label{eq:FCCondition3}
\end{aligned}
\end{eqnarray}
From the measurements at different rf amplitudes we find that the
parameter $a$ must be $a\sim10-100$ or larger in order for the
frequency comb technique to work reliably. This however is a
rather weak condition and is satisfied for a large class of
solid-state nuclear spin systems where $T_1\gg T_2$. This
demonstrates the wide applicability of the frequency comb
spectroscopy technique developed here.

%\bibliography{bibl}% Produces the bibliography via BibTeX.
%\end{document}

\end{document}